\documentclass{article}
\usepackage{spconf,amsmath,graphicx,xcolor}

\title{DATA AUGMENTATION FOR CHILDREN'S SPEECH RECOGNITION\\
THE ``ETHIOPIAN" SYSTEM FOR THE SLT 2021 CHILDREN SPEECH RECOGNITION CHALLENGE}
\name{
Guoguo Chen$^{*1}\thanks{* Equal contribution, listed here in alphabetical order.}$,
Xingyu Na$^{*2}$,
Yongqing Wang$^{*3}$,
Zhiyong Yan$^{*3}$,
Junbo Zhang$^{*3}$,
Sifan Ma$^3$,
Yujun Wang$^3$
}
\address{
$^1$Seasalt AI LLC, USA\\
$^2$Xiaoice, Microsoft Corporation, China\\
$^3$Xiaomi Corporation, China\\
{\small\tt guoguo@seasalt.ai, asr.naxingyu@gmail.com} \\
{\small\tt \{wangyongqing3, yanzhiyong, zhangjunbo1, masifan, wangyujun\}@xiaomi.com}
}

\begin{document}
\ninept
\maketitle
\begin{abstract}
  This paper presents the ``Ethiopian" system for the SLT 2021 Children Speech
  Recognition Challenge. Various data processing and augmentation techniques are
  proposed to tackle children's speech recognition problem, especially the lack
  of the children's speech recognition training data issue. Detailed experiments are
  designed and conducted to show the effectiveness of each technique, across
  different speech recognition toolkits and model architectures. Step by
  step, we explain how we come up with our final system, which provides
  the state-of-the-art results in the SLT 2021 Children Speech Recognition Challenge,
  with 21.66\% CER on the {\it Track 1} evaluation set ({\it 4th} place overall),
  and 16.53\% CER on the {\it Track 2} evaluation set ({\it 1st} place overall).
  Post-challenge analysis shows that our system actually achieves 18.82\% CER
  on the {\it Track 1} evaluation set, but we submitted the wrong version to
  the challenge organizer for {\it Track 1}.
\end{abstract}
\begin{keywords}
  speech recognition, children's speech, children's speech recognition, data
  processing, data augmentation, neural networks, deep learning
\end{keywords}
\section{Introduction}
Speech recognition has made tremendous progress in the past decade, thanks to
the rise of the deep learning techniques and the development of various
open-source speech recognition toolkits. To give an example, the original
Librispeech paper \cite{panayotov2015librispeech} in 2015 reports a
word-error-rate (WER) of 13.97\% for its {\it test-other} evaluation set, while
in \cite{zhang2020pushing}, the authors report 2.6\% for the same evaluation
set. That is a reduction of 81.4\% in terms of WER in just 5 years!

Most of the efforts, however, have been devoted to adults' speech recognition
systems. Children's speech recognition, on the other hand, remains very
challenging, despite the rising demands from applications such as smart
speakers and language learning.

There are a number of reasons why children's speech recognition remains
difficult. First, children typically have shorter vocal tracts and smaller vocal
folds, which lead to higher fundamental and formant frequencies. This means that
a large part of the children's speech spectrum will be ignored at the typical
speech recognition sampling rate of 16kHz \cite{gray2014child}. Second,
children's speech tend to have higher level of variability \cite{shivakumar2020transfer},
both acoustically and linguistically, which makes it harder to model. Finally,
unlike speech recognition for adults, where training data is generally available
and can be easily collected when necessary, collecting speech recognition
training data for children is oftentimes laborious and expensive. As a result,
children's speech recognition systems usually trained with limited amount of
children's speech data.

Researchers have proposed various techniques to alleviate the above problems.
In \cite{giuliani2003investigating,stemmer2003acoustic,shivakumar2014improving},
Vocal Tract Length Normalization (VTLN) is used to suppress the acoustic
variability introduced by children's shorter vocal tracts. In
\cite{gray2014child,shivakumar2014improving}, adaptation methods such as Maximum
Likelihood Linear Regression (MLLR), Maximum A-Posteriori (MAP) and Speaker
Adaptive Training (SAT) are found helpful when dealing with children's speech.
Those methods usually fall into the speaker normalization and adaptation
category, which handle the acoustic variability to some degree.
\cite{das1998improvements} tries to improve children's speech recognition from
the linguistic variability angle. It adopts language models trained on
children's speech, and finds that it yields better performance than language models
trained on adults' speech. \cite{shivakumar2020transfer} improves children's
speech recognition from the data insufficiency point of view. It adapts children's
model from adults' model using transfer learning, and finds it very helpful.

In this paper, we propose data augmentation for children's speech recognition.
Different data augmentation techniques, such as pitch perturbation, speech
perturbation, tempo perturbation, volume perturbation, reverberation
augmentation, spectral augmentation, are experimented with the SLT 2021 Children
Speech Recognition Challenge data. We conduct experiments with both Kaldi's
chain models \cite{Povey2016Purely} and ESPnet's attention-based encoder-decoder
end-to-end models \cite{Watanabe2018ESPnet}. We find that data augmentation
generally helps to improve model's robustness towards the acoustic variability
in children's speech, and as a result, improves children's speech recognition
performance across speech recognition toolkits and model architectures. Our
final submitted system is trained with ESPnet's attention-based encoder-decoder
end-to-end model, and it achieves 18.82\% CER and 16.53\% CER on
the challenge's {\it Track 1} and {\it Track 2} evaluation sets respectively.

The rest of this paper is organized as follows. In Section 2 we introduce the
SLT 2021 Children Speech Recognition Challenge. Section 3 presents our data
processing techniques, and Section 4 explains the data augmentation techniques
we propose for children's speech recognition. We describe our speech recognition
systems in Section 5, and experiments setup and results in Section 6. Finally,
Section 7 concludes the paper and discusses future work.

\section{The Challenge}
We briefly introduce the SLT 2021 Children Speech Recognition Challenge (CSRC)
in this section, including datasets and tracks.

\subsection{Datasets}
CSRC releases three datasets, namely {\it Set A}, {\it Set C1} and {\it Set C2}.
In addition to the three released datasets, external data listed on
OpenSLR \cite{openslr} is allowed for the
{\it Track 2} task.

\vspace{-2.0ex}\paragraph*{Set A.} The {\it Set A} dataset consists of 341.4
hours of Mandarin adult reading speech. It has 1999 speakers, with ages between
18 - 60 years old.
\vspace{-2.0ex}\paragraph*{Set C1.} The {\it Set C1} dataset consists of 28.6
hours of Mandarin child reading speech. It has 927 speakers, with ages between
7 - 11 years old.
\vspace{-2.0ex}\paragraph*{Set C2.} The {\it Set C2} dataset consists of 29.5
hours of Mandarin child conversational speech. It has 54 speakers, with ages
between 4 - 11 years old.
\vspace{-2.0ex}\paragraph*{External Data.} External data listed on
OpenSLR \cite{openslr} is allowed for the
{\it Track 2} task.

\subsection{Tracks}
CSRC has two separate tasks, namely the {\it Track 1} task and the {\it Track 2}
task. Teams can participate in either one of the tracks, or both tracks.
\vspace{-2.0ex}\paragraph*{Track 1.} In the {\it Track 1} task, only {\it Set A},
{\it Set C1} and {\it Set C2} can be used to train the acoustic and language
models.
\vspace{-2.0ex}\paragraph*{Track 2.} In the {\it Track 2} task, in addition to
the {\it Set A}, {\it Set C1} and {\it Set C2} datasets, external data listed on
OpenSLR \cite{openslr} can be used for acoustic
model training. Only the transcripts associated with the provided speech data
and the external speech data on OpenSLR \cite{openslr}
can be used for language model training.

\subsection{Evaluation}
Both tracks share the same evaluation dataset. The evaluation dataset consists
of 16.3 hours of Mandarin child reading and conversational speech, with
216 speakers. Character-error-rate (CER) is used to evaluate the model performance.

\section{Data Processing}
We split the official datasets into {\it train} and {\it dev} sets, and then
apply text normalization and tokenization for the {\it train} and {\it dev} sets
respectively before we train the speech recognition system.

\subsection{Data Partitioning}
Since CSRC focuses on children's speech recognition, our {\it dev} set is
selected only from the child portion of the datasets, that is {\it Set C1} and
{\it Set C2}. We randomly choose 20 speakers from {\it Set C1}, around 1.4 hours
of children's speech. We call this set {\it dev-011}. We also select 5 speakers
from {\it Set C2}, around 0.6 hours of children's speech. We call this set
{\it dev-018}. The rest of the released data is treated as our {\it train} set.

\subsection{Text Normalization}
The transcripts released by CSRC have been properly normalized. In addition to
the official text normalization, we also map symbols (e.g., ``$>$", ``@", etc.)
to their corresponding Chinese characters.

\subsection{Tokenization}
\begin{table}[ht]
\small
  \caption{Comparison of different tokenizers (CER)} 
  \vspace{0.5ex}
\centering 
\begin{tabular}{l c c } 
\hline\hline 
  Tokenization & dev-011      &  dev-018 \\ [0.5ex] 
  \hline
  mmseg        & 35.96        &63.15 \\
  jieba        & 35.52        &62.31 \\
\hline
\end{tabular}
\label{table:tokenization-comp} 
\end{table}

\begin{table*}[!t]
  \small
    \caption{List of all Kaldi data augmentation experiments (CER)} 
    \vspace{0.5ex}
  \centering 
  \begin{tabular}{l c c c} 
  \hline\hline 
    Kaldi-Exp                                                                & dev-011      & dev-018      & Average \\ [0.5ex] 
    \hline
    Exp1: Baseline                                                     & 10.2         & 22.36        & 16.28   \\
    Exp2: Baseline + 80-dim FBANK                                      & 10.22        & 21.95        & 16.09   \\
    Exp3: Baseline + 80-dim FBANK + SpecAug + depth tdnnf:17           & 9.95         & 21.66        & 15.81   \\
    Exp4: Exp3 + backward rnnlm: ngram order 4                         & 9.19         & 21.66        & 15.43   \\
    Exp5: Exp3 + forward rnnlm: ngram order 4                          & 9.12         & 20.98        & 15.05   \\
    Exp6: Exp3 + forward rnnlm: ngram order 5                          & 9.09         & 20.45        & 14.77   \\
    Exp7: Exp3 + 3-dim pitch features                                  & 9.92         & 21.62        & 15.77   \\
    Exp8: Exp3 + 3-dim pitch features + forward rnnlm: ngram order 5                                   & 9.16         & 21.04        & 15.1   \\
    Exp9: Exp6 + {\it Set A, C1, C2}:rp,\{sp,tp\}@\{0.85,0.9,1.1,1.15\}                                & 8.47         & 19.78        & 14.13   \\
    Exp10: Exp9 + remove clean + remove {\it Set A}:\{sp,tp\}@\{0.85,1.15\}                            & 8.63         & 19.26        & 13.95   \\
    Exp11: Exp10 + {\it Set A, C1, C2}:pp + {\it Set C1, C2}:\{sp,tp\}@\{0.88,1.12\}                   & 9.14         & 18.69        & 13.92   \\
    * Exp12: Exp9 + remove clean + {\it Set A, C1, C2}:pp + {\it Set C1, C2}:\{sp,tp\}@\{0.88,1.12\}     & 8.52         & 18.70        & 13.61   \\
  \hline
  \end{tabular}
  \begin{tabular}{l}
     * Post-challenge experiment ~~~~~~~~~~~~~~~~~~~~~~~~~~~~~~~~~~~~~~~~~~~~~~~~~~~~~~~~~~~~~~~~~~~~~~~~~~~~~~~~~~~~~~~~~~~~~~~~~~~~~~~~~~~~~~~~~~~~~~~~~~~~~~~~~~~~~~~~~~~~~~~~~~ \\
  \end{tabular}
  \label{table:kaldi-exp} 
  \end{table*}

The official training transcripts are not tokenized. We evaluate both mmseg
\cite{mmseg} and jieba \cite{jieba} tokenizers for this particular task. For
each tokenizer, we train a typical triphone system with the
Kaldi \cite{povey2011kaldi} toolkit, on our {\it train} set. We then evaluate
the recognition performance on our {\it dev-011} and {\it dev-018} sets. Table
\ref{table:tokenization-comp} reports the CER for both tokenizers on our {\it dev}
sets. It is clear that jieba outperforms mmseg on both {\it dev} sets. For the
rest of the task, jieba is used for tokenization.


\section{Data Augmentation}
Data augmentation has been proved to be very helpful to speech recognition
systems, especially when there is a training data mismatch, or when the training
data is insufficient \cite{ko2015audio,peddinti2015jhu,peddinti2015reverberation,Qian2017Mismatched}.
CSRC's evaluation data only consists of child reading and conversational speech. Its
released training data, on the other hand, is mostly adult speech, with around
340 hours of adult reading speech, plus around 30 hours of child reading speech
and another 30 hours of child conversational speech. Given the limited amount
of child speech in the training data, we believe data augmentation can play a
big role in this particular task.

\subsection{Pitch Perturbation}
We propose to use pitch perturbation for adult speech to make it closer to child
speech, effectively increasing the training data for child speech. We use SoX's
``pitch" option for pitch perturbation, which shifts the original speech's pitch
by ``cents", i.e., 1/100th of a semitone. We experiment with different shift
values, and find shifting adult speech's pitch up by 250 - 370 cents yields the
best performance. For each utterance in the adult speech ({\it Set A}) dataset,
we randomly pick a value between 250 - 370, and shift the utterance's pitch up
by that value. Note that we only intend to apply pitch perturbation to adult
speech. But in our experiments, we made a mistake and applied it to the whole
dataset.

\subsection{Speed Perturbation}
Speed perturbation generally improves speech recognition
performance \cite{ko2015audio}. Speed perturbation is usually performed by
resampling the original speech signal. As a result, speed perturbation affects
both pitch and tempo. For adult speech, when we increase the original speed to
a faster rate, it also pushes the pitch higher, making it closer to child speech.
So technically, children's speech recognition should benefit more from speed
perturbation. We use SoX's ``speed" option for speed perturbation.
For adult speech ({\it Set A}), we create 2 additional copies
of the original data, at 90\% and 110\% of the original speed. For child speech
({\it Set C1} and {\it Set C2}), we create 6 additional copies of the original
data, at 85\%, 88\%, 90\%, 110\% 112\% and 115\% of the original speed
respectively.

\subsection{Tempo Perturbation}
Tempo perturbation modifies the speaking rate of the original speech signal without
changing its pitch \cite{kanda2013elastic}. By adding additional training data
at various speaking rate, it makes the speech recognition model robust to
different speakers. We use SoX's ``tempo" option for tempo perturbation, which
internally implements the waveform-similarity-based synchronized overlap-add (WSOLA)
algorithm \cite{verhelst1993overlap}.
For adult speech ({\it Set A}), we create 2 additional copies
of the original data, at 90\% and 110\% of the original tempo. For child speech
({\it Set C1} and {\it Set C2}), we create 6 additional copies of the original
data, at 85\%, 88\%, 90\%, 110\% 112\% and 115\% of the original tempo respectively.

\subsection{Volume Perturbation}
Volume perturbation is a simple but effective technique to improve speech
recognition model's robustness towards audio volume variations. We use SoX's
``vol" option for volume perturbation. For all the utterances in our training
data, including those created from other data augmentation techniques, we
perturb the audio volume by a factor between 0.125 and 2.

\subsection{Reverberation Perturbation}
Reverberation perturbation is critical for far-field speech recognition
\cite{peddinti2015jhu}. We do not expect to see a lot of reverberant speech
in the final evaluation set, but we add a small portion of reverberation perturbed
data anyways to increase the model robustness. We use SoX's ``reverb" option
for reverberation perturbation. We apply reverberation perturbation to all the
three datasets {\it Set A}, {\it Set C1} and {\it Set C2}.

\subsection{Spectral Augmentation}
Spectral augmentation is typically applied directly to the feature inputs of
a neural network \cite{park2019specaugment}. The augmentation policy consists of
warping the features, masking blocks of frequency channels, and masking blocks
of time steps. We use spectral augmentation in all our systems, which does
improve the performance in this task.

\section{System Description}
We use Kaldi \cite{povey2011kaldi} for our early experiments, e.g., figuring out
what data augmentation techniques to use, and ESPnet \cite{Watanabe2018ESPnet}
for our final submission.

\subsection{Kaldi System}

Early experiments are based on Kaldi's {\it multi-cn} recipe, which implements
a typical chain model \cite{Povey2016Purely}.
First, a GMM-HMM model is trained to obtain the alignments.
Second, data augmentation techniques described in previous sections,
such as volume and speed perturbation, are applied.
I-vectors \cite{dehak2010front} are extracted and pasted to the basic acoustic features,
as most Kaldi's recipes \cite{peddinti2015reverberation} do.
Finally, a neural network is trained with both cross-entropy and LF-MMI criteria.

Some tiny modifications are made on top of the original {\it multi-cn} recipe.
We use 80-dimensional FBANK as the basic acoustic features,
while the default feature dimension is 40.
We increase the number of factorized time-delay neural network (TDNN-F) layers to 17,
while the default neural network layout of the {\it multi-cn} recipe stacks 6
convolutional neural network (CNN) layers and 12 TDNN-F layers.
The re-segmentation step of the original {\it multi-cn} recipe is also removed
to speed up the overall training.

Other data augmentation strategies are also applied to the recipe.
First, spectral augmentation is performed by adding Kaldi's built-in nnet3 component
{\it spec-augment-layer} during the neural network training.
Second, for speed perturbation, besides the default speed factors (0.9 and 1.1),
we use additional factors 0.8 and 1.15.
Third, tempo perturbation is added with the same factors as the speed perturbation.
And finally, reverberation perturbation is also performed, as described in previous sections.

As for the language models, we add Recurrent Neural Network Language Model (RNN LM) 
for rescoring to further improve the performance.

\subsection{ESPnet System}
The end-to-end flavor system is built using ESPnet \cite{Watanabe2018ESPnet}. The data
augmentation strategies experimented in Kaldi are reused. Convolution-augmented
Transformer (Conformer) is applied with relative positional encoding-based self
attention \cite{conformer}. Encoder is constructed using 12 layers of Conformer
blocks. Each block consists of a feed-forward module, a multi-head self
attention (MHSA) module and a convolution module followed by another feed-forward
module. The hidden dimension of the linear layers in feed-forward modules is
2048. The output dimension of each blocks and the dimension of the MHSA are both
256. Specifically, the number of heads in MHSA is 4. The kernel size of the
convolution module is 15.

The input to the encoder goes through a SpecAugment \cite{park2019specaugment} layer followed by
a 2-dimensional convolution block with 1/4 subsampling. The block is a stack of
2 2-dimensional convolutions with kernel size 3 and subsampling rate 2, each of
which are followed by ReLU activations. Each of the blocks in the encoder begins with
a layer-wise normalization.

The decoder is a stack of 6 decoder blocks. Each block consists of 2 MHSA layers
and 1 feed-forward layer. The dimension of the MHSA layer is 256, and the hidden dimension
of the feed-forward layer is 2048. Each layer begins with a layer-wise normalization. The
target of the decoder is represented using embedding of 256. The activation of
the decoder goes through another layer-wise normalization and linear transform
before output.

In both the encoder and the decoder, the dropout rate of attention components is 0,
while that of the other components are 0.1. The training loss is default to
ESPnet \cite{Watanabe2018ESPnet} with tuning factor set to 0.3 and label
smoothing weight set to 0.1.

The language model (LM) used in this system is token-based 4-gram. For evaluation, hybrid
CTC/attention decoding is adopted with CTC weight 0.5 \cite{decoding}. Besides
CTC, ngram score is used with weight 0.5.

\subsection{Final {\it Track 1} System}
\begin{table}[ht]
\small
  \caption{Final {\it Track 1} System: breakdown of data augmentation} 
  \vspace{0.5ex}
\centering 
\begin{tabular}{r r } 
\hline\hline 
  Data Augmentation & Hours \\ [0.5ex] 
  \hline
  {\it Set A} *                                                                               & 341.8 \\
  {\it Set C1, C2} *                                                                          & 55.1 \\
  \hline
  {\it Set A, C1, C2} + rp + vp                                                               & 396.9 \\
  {\it Set A, C1, C2} + pp + vp                                                               & 396.9 \\
  {\it Set A} + sp@\{0.9,1.1\} + vp                                                           & 690.6 \\
  {\it Set A} + tp@\{0.9,1.1\} + vp                                                           & 690.6 \\
  {\it Set C1, C2} + sp@\{0.85,0.88,0.9,1.1,1.12,1.15\} + vp                                  & 342.7 \\
  {\it Set C1, C2} + tp@\{0.85,0.88,0.9,1.1,1.12,1.15\} + vp                                  & 342.7 \\
  \hline
  Total                                                                                       & 3257.3 \\
\hline
\end{tabular}
\begin{tabular}{l l}
   pp :& Pitch perturbation\\
   rp :& Reverberation perturbation \\
   sp :& Speed perturbation, values inside the curly braces are \\
       & different perturbation parameters \\
   tp :& Tempo perturbation, values inside the curly braces are \\
       & different perturbation parameters \\
   vp :& Volume perturbation \\
   *   & Our number is a little bit less than the official number \\
\end{tabular}
\label{table:track1-data-breakdown} 
\end{table}

While the effectiveness of the data augmentation techniques has been proved in Kaldi-based
experiments, the final system for both {\it Track 1} and {\it Track 2} are created using
ESPnet. In {\it Track 1}, a warm up model is trained using data section 1, 2 and 5 in
table \ref{table:track1-data-breakdown} for 50 epochs with learning rate
1.0. The average model of last 10 epochs is saved as the base model. After that, the
model is fine-tuned using all the data listed in
table \ref{table:track1-data-breakdown} with learning rate 0.25. The final model
is the average of the 5 best models in the fine-tuning stage w.r.t. the loss on
the validation set. In the decoder, the number of tokens the for {\it Track 1} task is 5784.

\subsection{Final {\it Track 2} System}
\begin{table}[ht]
\small
  \caption{Final {\it Track 2} System: breakdown of data augmentation} 
  \vspace{0.5ex}
\centering 
\begin{tabular}{r r } 
\hline\hline 
  Data Augmentation & Hours \\ [0.5ex] 
  \hline
  {\it Set A}                                                                                 & 341.8 \\
  {\it Set C1, C2}                                                                            & 55.1 \\
  OpenSLR *                                                                                 & 1374.3 \\
  \hline
  {\it Set A, C1, C2} + rp + vp                                                               & 396.9 \\
  {\it Set A, C1, C2} + pp + vp                                                               & 396.9 \\
  {\it Set A} + sp@\{0.9,1.1\} + vp                                                           & 690.6 \\
  {\it Set A} + tp@\{0.9,1.1\} + vp                                                           & 690.6 \\
  {\it Set C1, C2} + sp@\{0.85,0.88,0.9,1.1,1.12,1.15\} + vp                                  & 342.7 \\
  {\it Set C1, C2} + tp@\{0.85,0.88,0.9,1.1,1.12,1.15\} + vp                                  & 342.7 \\
  OpenSLR + sp@\{0.9,1.1\} + vp                                                               & 2776.4 \\
  \hline
  Total                                                                                       & 7408 \\
\hline
\end{tabular}
\begin{tabular}{l l}
   pp :& Pitch perturbation\\
   rp :& Reverberation perturbation \\
   sp :& Speed perturbation, values inside the curly braces are \\
       & different perturbation parameters \\
   tp :& Tempo perturbation, values inside the curly braces are \\
       & different perturbation parameters \\
   vp :& Volume perturbation \\
   *   & Datasets SLR-\{18, 33, 38, 47, 62, 68\} are used \\
\end{tabular}
\label{table:track2-data-breakdown} 
\end{table}

Due to the increased amount of training data in the {\it Track 2} task, the
number of tokens in the decoder is increased to 6491. The
number of blocks in the encoder and that in the decoder are increased to 18 and 9 respectively, and the
attention width is also increased to 512 with 8 multi-heads. The {\it Track 2} model is
trained from scratch using all the data in table
\ref{table:track2-data-breakdown}. Instead of checkpointing based on epochs,
models are saved every 10000 batches. For the final submission, 8
best models w.r.t. the loss on validation set are averaged as the final output model.

\section{Results and Analysis}

\subsection{Kaldi Experiments}

All data augmentation experiments are conducted with the Kaldi toolkit, as shown
in Table \ref{table:kaldi-exp}. We use Kaldi's {\it multi-cn} recipe as our
baseline, which gives 16.28\% CER on our {\it dev} set. By increasing the feature
dimension from 40 to 80, we reduce the CER to 16.09\%, and by adding spectral
augmentation and additional network layers, we further reduce it to 15.81\%. From
this point, various language models are experimented. We find that a forward
Recurrent Neural Network Language Model (RNN LM) with 5-gram rescoring order (Exp6) gives
the best result, which bring the CER down to 14.77\%. Pitch features are also
added to the Exp3 experiment setting, but they do not seem to improve the
performance. We therefore do not use pitch features in the rest of the experiments.
On top of Exp6, we add speed and tempo perturbation with parameters 0.85, 0.9, 1.1
and 1.15. We also add reverberation perturbation for {\it Set A}, {\it Set C1}
and {\it Set C2} (Exp9), which reduces the CER from 14.77\% to 14.13\%. We are a little
bit concerned that our system is overfitting to the adult speech, so we remove
half of the speed and tempo perturbation (with parameters 0.85 and 1.15) data for
the adult speech set {\it Set A}. We also remove the data re-segmentation and
cleaning step as that is becoming very slow. With those two changes, CER improves
to 13.95\& (Exp10). From there, we add pitch perturbation for all three datasets
{\it Set A}, {\it Set C1} and {\it Set C2}, and we also add more speed and tempo
perturbation for the child speech datasets {\it Set C1} and {\it Set C2}. That
brings the CER to 13.92\% (Exp11). Remember in Exp10 we remove half of the speed
and tempo perturbation data for the adult speech set {\it Set A}. If we keep the
same perturbation for adult speech, and in addition add pitch perturbation and
more speech and tempo perturbation for the child speech set, we are able to
achieve 13.61\% CER on our {\it dev} set.

\subsection{ESPnet Experiments}
\begin{table}[ht]
\small
  \caption{Kaldi and ESPnet Comparison (CER)} 
  \vspace{0.5ex}
\centering 
\begin{tabular}{l c c c} 
\hline\hline 
  Systems              & dev-011      & dev-018      & Average \\ [0.5ex] 
  \hline
  Kaldi-Exp3           & 9.95         & 21.66        & 15.81   \\
  ESPnet               & 8.3          & 21.60        & 14.95   \\
\hline
\end{tabular}
\label{table:espnet-exp} 
\end{table}
We also run experiments through ESPnet's attention-based encoder-decoder
end-to-end models. Table \ref{table:espnet-exp} shows a comparison between ESPnet
and Kaldi with the same amount of augmented data (Exp3 augmentation setting).
On our {\it dev} set, ESPnet's
end-to-end model slightly outperforms Kaldi's chain model.

\subsection{Final Submission}
\begin{table}[ht]
\small
  \caption{Final Submission (CER)} 
  \vspace{0.5ex}
\centering 
\begin{tabular}{l c c c} 
\hline\hline 
  Track                        & evaluation  \\ [0.5ex] 
  \hline
  ESPnet-track1                & 18.82  \\
  ESPnet-track2                & 16.48  \\
\hline
\end{tabular}
\label{table:final-exp} 
\end{table}
From the previous section we learn that ESPnet's end-to-end model slightly outperforms
Kaldi's chain model. We therefore use ESPnet for our final submission. Our best
Kaldi system before the submission is Exp11, so we use Exp11's augmentation
settings, and run it through ESPnet's end-to-end model. Results on the final
evaluation set are shown in Table \ref{table:final-exp}. It is worth mentioning
that Exp11 is not our best setting. Post-challenge experiments reveal that Exp12,
which has more augmentation data, outperforms Exp11 with a healthy margin on
Kaldi's chain model. We expect the Exp12 setting will further our CERs on the
evaluation set.

\section{Conclusions}
We have demonstrated that data augmentation is simple and effective for children's
speech recognition. Given the short time window for development, we believe
we have not fully utilized the power of data augmentation. Further experiments
are encouraged.

\clearpage
\bibliographystyle{IEEEbib}
\bibliography{strings,refs}

\end{document}